\documentclass[11pt]{article}
\usepackage[dvips]{graphicx}
\usepackage{latexsym}
\usepackage{euscript}

\newcommand{\msbar}{\overline{\mbox{\scriptsize MS}}}
\newcommand{\MSbar}{\overline{\mbox{MS}}}
\newcommand{\RI}{\mbox{\scriptsize RI}}
\newcommand{\ri}{\mbox{\scriptsize RI}}
\newcommand{\RIb}{\mbox{MOM}}

\newcommand{\Pj}{\mbox{I}\!\!\mbox{P}}

\newcommand{\Tr}{\mbox{Tr}\;} 
\newcommand{\<}{\langle}
\renewcommand{\>}{\rangle}
\newcommand{\bc}{\begin{center}}
\newcommand{\ec}{\end{center}}
\newcommand{\be}{\begin{equation}}
\newcommand{\ee}{\end{equation}}
\newcommand{\bea}{\begin{eqnarray}}
\newcommand{\eea}{\end{eqnarray}}
\newcommand{\ba}{\begin{eqnarray}}
\newcommand{\ea}{\end{eqnarray}}
\newcommand{\brr}{\begin{array}}
\newcommand{\err}{\end{array}}

\newcommand{\pslash}{p\!\!\!/\,}

\newcommand{\hatk}{\hat{k}_\mu}
\newcommand{\hatkk}{\hat{k}^2}
\newcommand{\bark}{\bar{k}_\mu}
\newcommand{\barkk}{\bar{k}^2}

\newcommand{\hatpp}{\hat{p}^2}

\newcommand{\barpp}{\bar{p}^2}

\def\dfrac#1#2{{\displaystyle {#1 \over #2}}}

\newcommand{\nn}{\nonumber}
\newcommand{\id}{\hbox{1$\!\!$1}}

\newcommand{\simge}{\ \lower-
1.2pt\vbox{\hbox{\rlap{$>$}\lower5pt
\vbox{\hbox{$\sim$}}}}\ }

\begin{document}
\pagestyle{empty} 
\vspace{-0.6in}
\begin{flushright}
BUHEP-00-13\\
MIT-CTP-2991\\
July 2000
\end{flushright}
\vskip 1.5in

\centerline{\large {\bf{Perturbative Renormalization of Weak-Hamiltonian}}} 
\centerline{\large {\bf{Four-Fermion Operators with Overlap Fermions}}}
\vskip 0.6cm
\centerline{\bf Stefano Capitani$^{1}$, Leonardo Giusti$^{2}$}
\vskip 0.5cm
\centerline{$^1$ Massachusetts Institute of Technology}
\centerline{Center for Theoretical Physics, Laboratory for Nuclear Science}
\centerline{77 Massachusetts Avenue, Cambridge MA 02139 USA.}
\centerline{e-mail: stefano@mitlns.mit.edu}
\vskip 0.2cm
\centerline{$^2$ Boston University - Department of Physics} 
\centerline{590 Commonwealth Avenue, Boston MA 02215 USA.}
\centerline{e-mail: lgiusti@bu.edu}

\vskip 1.0in
\begin{abstract}
The renormalization of the most general dimension-six 
four-fermion operators without power subtractions is studied
at one loop in lattice perturbation theory using overlap
fermions. As expected, operators with different chirality
do not mix among themselves and parity-conserving and parity-violating
multiplets renormalize in the same way. The renormalization constants 
of unimproved and improved operators are also the same. 
These mixing factors are necessary to determine physical 
matrix elements relevant 
to many phenomenological applications of weak interactions. 
The most important are the $K^0-\bar{K}^0$ and $B^0-\bar{B}^0$ 
mixings in the Standard Model and beyond, the $\Delta I =1/2$ rule and 
$\epsilon^\prime/\epsilon$.
\end{abstract}
\vfill
\pagestyle{empty}\clearpage
\setcounter{page}{1}
\pagestyle{plain}
\newpage 
\pagestyle{plain} \setcounter{page}{1}

\newpage

\section{Introduction}
Since the original proposals of using lattice QCD to study hadronic 
weak decays~\cite{CMP}, 
substantial theoretical and numerical progress has been made. 
In the most popular lattice regularizations, i.e. Wilson and staggered
fermions, the main theoretical aspects of the renormalization of 
composite four-fermion operators are fully understood~\cite{boc,shape1}. 
The calculations of the matrix elements relevant for the 
$K^0$--$\bar K^0$ and $B^0$--$\bar B^0$ mixings have reached a level of 
accuracy which is unmatched by any other approach~\cite{ksgp,Sharpe96}; 
increasing precision has also been achieved in determining the
electro-weak penguin amplitudes necessary for the prediction
of the CP-violation parameter $\epsilon^\prime/\epsilon$ 
\cite{bs}-\cite{NoiDELTAS=2}.
Finally, matrix  elements of  $\Delta S=2$ operators, relevant 
to the study of FCNC effects in several extensions of the Standard Model
(supersymmetry, left-right symmetric models, multi-Higgs models
etc.), have also been computed~\cite{NoiDELTAS=2_old}-\cite{ds2susy}. 
Some of these lattice predictions have been fundamental in constraining the 
parameters of the CKM matrix in the Standard Model (SM) 
\cite{CKM,ep/e} and beyond \cite{CKM_beyond}.\\ 
\indent Nevertheless, some fundamental phenomena, such as the $\Delta I=1/2$ rule 
in non-leptonic kaon decays, and the value of
$\epsilon^\prime/\epsilon$, which measures the direct CP violation in 
kaon decays, are far from being understood. 
In the Standard Model these effects can be explained only if the non-perturbative 
physics gives contributions
to matrix elements definitively larger than their factorized values
\cite{ep/e}. Therefore a non-perturbative determination of the relevant 
matrix elements is crucial for predicting these quantities.
Lattice QCD is the only method which can address these problems 
from first principles. Techniques have been developed
for both Wilson and staggered fermions, but these methods have not yet
produced useful results \cite{bs,old_lat}.\\ 
\indent To compute non-leptonic weak matrix elements it is essential 
to construct renormalized operators in definite chiral representations.
In the Wilson or SW-Clover lattice regularizations, bare operators do not
have a definite chiral behavior due to the presence of the symmetry
breaking term in the action. Renormalized operators with the correct chiral 
properties are recovered as linear combinations of operators with different chirality. 
The Standard Model $O^{\Delta S =2}$ operator is the most popular example:
its matrix element between pseudoscalar states should vanish as
$M_\pi^2$ in the limit $M_\pi \rightarrow 0 $, while the matrix elements
of the wrong-chirality operators which mix with it 
are expected to go to a constant (and in fact in the 
kaon mass region they are 2 to 10 times larger than the 
SM one). Therefore the correct chiral behavior of
$O^{\Delta S =2}$ is recovered only if the 
finite mixing constants are known with high precision.
This has been a long-standing problem and has only been solved 
using Non-Perturbative (NP) renormalization techniques \cite{BK_chiral,japanbk,tassos}.
The situation becomes even worse when the lack of chiral symmetry 
complicates mixings with lower dimensional operators. This is one of the 
obstacles to a reliable computation of the matrix elements 
relevant for $\Delta I =1/2$ and $\epsilon^\prime/\epsilon$.
On the other hand, with staggered fermions chiral symmetry is 
preserved, but solving the doubling problem and defining operators 
with the correct flavor and spin quantum numbers is far from trivial.  

Only recently has it been understood 
\cite{hasenfratz}-\cite{luscher}
that chiral and flavor symmetries  can be preserved 
simultaneously on the lattice, without 
fermion doubling, if the fermionic operator $D$ satisfies the 
Ginsparg-Wilson Relation (GWR) \cite{GW}:
\be\label{eq:GW}
\gamma_5 D + D \gamma_5 = \frac{a}{\rho} D \gamma_5 D \; .
\ee
The GWR implies an exact symmetry of the fermion 
action at non zero-lattice spacing, which may be regarded as a 
lattice form of the standard chiral rotation \cite{luscher}.
Nevertheless, it is important to stress that locality, the absence of 
doubler modes and the correct classical continuum limit are not 
guaranteed by the GWR in eq.~(\ref{eq:GW}). Indeed, there exist 
lattice fermion actions which satisfy the GWR but which
do not meet the above requirements \cite{chiu}.
A breakthrough in this field was achieved through the domain-wall
formulation of lattice fermions \cite{walls} and 
by Neuberger through the overlap formulation \cite{neub1}. He
found a solution  of the GWR which satisfies all the 
above requirements and is local\footnote{From eq.~(\ref{eq:opneub}) it is
clear that the overlap operator is not ultra-local. The Neuberger 
kernel satisfies a more general definition of locality, i.e. it 
is exponentially suppressed at large distances with a decay rate 
proportional to $1/a$.} \cite{pilar}:
\ba\label{eq:opneub}
D_N &=& \frac{\rho}{a}\left( 1 + X\frac{1}{\sqrt{X^\dagger X}}\right)\nonumber\\
X &=& D_W -\frac{1}{a}\rho\; ,
\ea
where $D_W$ is the Wilson-Dirac operator and $0<\rho< 2r$ (see below). 
A further remarkable property of Ginsparg-Wilson fermions 
is the absence of $O(a)$ discretization errors in the action and 
therefore in the spectrum of the theory. Nevertheless, 
the local fermionic operators have to be improved
to remove the $O(a)$ effects in matrix elements. 
This step is greatly simplified by the Ginsparg-Wilson 
relation, as it allows the construction 
of $O(a)$-improved operators to all orders 
in $g^2$ \cite{qcdsf}, which renormalize with the same renormalization 
factors of the unimproved
ones. On the other hand, the complicated form of the Neuberger operator in 
eq.~(\ref{eq:opneub}) renders its numerical implementation quite demanding
for the present generation of computers. However, some progress has been achieved 
\cite{simulazioni} and Monte Carlo simulations seem to be already 
feasible, at least in the quenched approximation.

Once the action satisfies the GWR and all the properties described
above, the quark masses renormalize only multiplicatively and  
mixings of operators with different chirality are forbidden
even at finite cut-off \cite{hasenfratz2}. The most general set of dimension-six 
four-fermion operators without power subtractions is
\ba\label{eq:basstart2}
O_{\Gamma_A \Gamma_B} & = & 
\bar \psi_1\Gamma^{l}_A \psi_2 \cdot \bar \psi_3\Gamma^{l}_B \psi_4\\
O^F_{\Gamma_A \Gamma_B} & = & 
\bar \psi_1\Gamma^{l}_A \psi_4 \cdot \bar \psi_3\Gamma^{l}_B \psi_2\; .\nonumber
\ea
The main result of this paper is the evaluation of the mixing pattern of these operators
at one loop in perturbation theory in the overlap lattice regularization: we 
compute their renormalization constants and we show how the Neuberger action 
greatly simplifies the mixing among them.

The renormalization constants are necessary for extracting 
physical matrix elements from numerical simulations.
The matrix elements of the operators studied in this work are
relevant in many phenomenological applications of weak interactions.
They are necessary for predicting 
the $K^0-\bar{K}^0$ and $B^0-\bar{B}^0$ mixing amplitudes in the Standard Model 
and in several extensions of it (supersymmetry, left-right symmetric 
models, multi-Higgs models etc.). By using chiral perturbation 
theory, they can also be related to the $\Delta I =3/2$ matrix elements 
relevant for the prediction of $\epsilon'/\epsilon$. 
They are also necessary
for estimating the $B^0_s-\bar{B}^0_s$ width difference and 
the $O(1/m_b^3)$ corrections in inclusive b-hadron decay 
rates \cite{beneke}. The renormalization constants we have computed are also 
relevant for the study of the $\Delta I=1/2$ rule and 
$\epsilon'/\epsilon$ on the lattice. The operators involved 
in these computations can mix with same- 
as well as lower-dimension operators. The key 
observation in these cases \cite{testa} is that the lower-dimension
operators do not change the anomalous dimensions of the 
original operators. Therefore the mixings among dimension-six
operators can be obtained from those of the operators studied in this paper.

Perturbative computations with the overlap-Dirac operator are much more 
cumbersome than for Wilson fermions, due to the complicated
structure of the vertices in the action (see Appendix \ref{appb}). For this
reason it is very useful to use an intermediate renormalization scheme 
to separate as much as possible the computations 
performed using lattice perturbation theory (where, for example, the 
Fierz rearrangements work) from the continuum perturbation theory which 
is much simpler. The RI renormalization scheme proposed in 
\cite{RI,NP} is the optimum choice:
we first renormalize lattice operators in the RI scheme
using only lattice perturbation theory and then the matching  
with one of the $\MSbar$ schemes is done in continuum
perturbation theory. Lattice and continuum regularizations  
are used independently taking full advantage 
of their properties.\\ 
\indent In Ref.~\cite{vicari}, renormalization constants of local 
bilinear fermion operators have been computed in the $\MSbar$ renormalization
scheme using the lattice overlap regularization. We computed 
these renormalization constants in the RI scheme, matched
them in $\MSbar$ and verified that our results are in agreement 
with those of Ref.~\cite{vicari}. Then we computed the renormalization constants 
of the four-fermion operators we are interested in following two independent
ways: the direct computation, like in Ref.~\cite{stefano}, and one which uses
the Fierz and charge conjugation rearrangements to
compute the four-fermion Green's functions in terms of the bilinear ones
\cite{guido,sharpe}. The agreement of the results is also a non-trivial check that 
the numerical integrals with the Neuberger propagators and vertices are well
estimated. 

A further determination of the renormalization constants considered in
this paper can be obtained using  numerical NP methods 
such as those in Refs.~\cite{NP,luscher_np}, applied to Neuberger fermions. However,
perturbative estimates are very useful because often they are very good
approximations and they furnish a consistency check of the NP methods. Moreover,
for Neuberger fermions the perturbative computations can remain 
the only determinations of the renormalization constants for some time.    

The paper is organized as follows: in section \ref{sec:definitions} we define 
the overlap fermion action used; in section \ref{sec:bils} we discuss
the bilinear renormalization constants; in section 
\ref{sec:4-fermions} we address the mixing of the four-fermion operators 
and report the results for the renormalization
constants; in section \ref{sec:conc} we state
our conclusions. 

\section{Basic definitions}\label{sec:definitions} 
The QCD lattice regularization we consider for massless
fermions is described by the action 
\be   
S_L = \frac{6}{g_0^2}\sum_{P} \bigg[ 1 - \frac{1}{6}   
\Tr \bigl[ U_P + U_P^{\dagger} \bigr] \bigg]  + 
\sum_{i=1}^{N_f}\sum_{x,y} \bar{\psi}_i(x) \, 
D_{N}(x,y) \, \psi_i(y),
\label{eq:sg}   
\ee   
where, in standard notation, $U_P$ is the Wilson plaquette 
and $g_0$ the bare coupling constant. 
$D_{N}$ is the Neuberger-Dirac operator defined 
in eq.~(\ref{eq:opneub}), where the massless Wilson operator $D_W$ is defined as 
\be\label{eq:WDO}
D_W = \frac{1}{2} \gamma_\mu (\nabla_\mu + \nabla^*_\mu)
-\frac{r}{2} a \nabla^*_\mu\nabla_\mu
\ee
and $\nabla_\mu$ and $\nabla^*_\mu$ are the forward and backward lattice
derivatives, i.e.
\ba
\nabla_\mu \psi(x) & = & \frac{1}{a}\Big[U_\mu(x) \psi(x + a \hat \mu) -
\psi(x)\Big]\nonumber\\
\nabla_\mu^* \psi(x) & = & 
\frac{1}{a}\Big[ \psi(x) - U^{\dagger}_\mu(x-a \hat \mu) 
\psi(x-a\hat\mu) \Big]\; .\label{eq:derivative}
\ea
The range of the Wilson parameter is $0<r<1$ and $U_\mu(x)$ is the lattice 
gauge link. Eq.~(\ref{eq:GW}) implies a continuous
symmetry of the fermion action in (\ref{eq:sg}), which may interpreted as a 
lattice form of chiral symmetry \cite{luscher}:
\be\label{eq:luscher}
\delta \psi  = \gamma_5 \left(1-\frac{1}{2\rho}a D_{N}\right)\psi, \qquad 
\delta \bar\psi  = \bar \psi \left(1-\frac{1}{2\rho}a D_{N}\right) \gamma_5 . 
\ee
The corresponding flavor non-singlet chiral transformations are defined
including a color group generator in eq.~(\ref{eq:luscher}). 
The generalization to massive fermions is simple \cite{nied}: in  
eq.~(\ref{eq:sg}) $D_{N}$ has to be replaced by
\be
D_{N} \longrightarrow \left(1-\frac{1}{2\rho} a m^i_0\right)D_{N} + m^i_0
\ee
where $m^i_0$ is the bare physical quark mass of flavor  $i$. 
The Feynman rules of the action defined in Eq.~(\ref{eq:sg})
are given in Appendix \ref{appb} and are in agreement with those 
used by \cite{vicari,japan,stefano_neub}. Throughout this 
paper we will use only mass independent renormalization schemes (RI and 
$\MSbar$) and therefore all our computations are performed with massless 
quarks. 

\section{Renormalization of Bilinear Operators}
\label{sec:bils}
In this section we set our notation for bilinear operators and we 
report the results we have obtained for their renormalization constants 
in the RI and $\MSbar$ schemes.\\
\indent A generic non-singlet quark bilinear is defined as
\be
O_\Gamma(x)=\bar\psi_1(x)\Gamma\psi_2(x)
\ee
where the flavors $\psi_1-\psi_2$ are different and $\Gamma$ is a generic 
Dirac matrix. The GWR ensures that no lattice artifacts of $O(a)$
are present in the action and therefore also not in the spectrum of the theory. 
However, matrix elements of operators 
are still affected by $O(a)$ discretization errors that 
have to be removed by improving the fermionic operators.
In Refs.~\cite{qcdsf} it is shown that, for massless quarks,
the improved bilinear operator is given by
\begin{equation}
O_\Gamma^I = \bar{\psi} \Big(1-\frac{1}{2\rho} a D_N\Big) \, \Gamma \, 
\Big(1-\frac{1}{2\rho} a D_N\Big) \psi \; 
\label{eq:improvedneubergerbil}\; .
\end{equation}
It can be proven \cite{vicari,qcdsf} that the renormalization constants of 
$O_\Gamma^I$ are the same as those of the corresponding 
$O_\Gamma$. Therefore all the results obtained in this section for local
bilinear operators also remain valid for the corresponding improved 
operators.\\
\indent The renormalized bilinear operators are defined by   
\be   
\widehat O_\Gamma(\mu) = Z_\Gamma(a\mu) O_\Gamma(a)   \; 
\label{eq:zodef}   
\ee   
and the chiral symmetry in eq.~(\ref{eq:luscher}) imposes the 
constraints 
\ba
Z_A & = & Z_V \nonumber\\
Z_P & = & Z_S\; 
\ea
on the renormalization constants\footnote{If the conserved axial and vector currents corresponding
to the chiral symmetry in eq.~(\ref{eq:luscher}) were used, then
$Z_V=Z_A=1$ \cite{hasenfratz}.}.

The quark propagator in momentum space is denoted by $S(p)$
(for the conventions adopted throughout the paper see 
Appendix \ref{appa}). 
The two-point fermionic Green's function of a bilinear 
inserted at the origin ($x=0$) is  
\be   
G_\Gamma(x_1,x_2) = \langle \psi_1(x_1) O_\Gamma(0) \bar \psi_2(x_2) \rangle
\; ,  
\label{eq:gos}   
\ee
its Fourier transform with equal external momenta is
\be
G_\Gamma(p) = \int dx_1 dx_2 e^{-ip(x_1 - x_2)} G_\Gamma(x_1,x_2)\; ,   
\label{eq:gp}   
\ee
and the corresponding amputated correlation function is defined as
\be
\Lambda_\Gamma(p) = S^{-1}(p) G_\Gamma(p) S^{-1}(p)\; .
\ee 
The renormalized quark propagator is 
\be   
\widehat S (p,\mu) =    
Z_\psi(a \mu) S (p,a)   \; ,
\label{eq:cp}   
\ee   
where $Z_\psi^{1/2}$ is the quark field renormalization constant, and 
the renormalized Green's functions are    
\ba
\widehat G_\Gamma(p,\mu) & = &
Z_\psi(\mu a) Z_\Gamma(\mu a) G_\Gamma (p,a)  \; ,\\   
\widehat \Lambda_\Gamma(p,\mu) & = & 
Z^{-1}_\psi(\mu a) Z_\Gamma(\mu a) 
\Lambda_\Gamma (p,a)  \; .  
\label{eq:clbil}   
\ea  

The anomalous dimensions of composite operators 
are defined as 
\be
\gamma_\Gamma (\alpha_s) = - 2Z_\Gamma^{-1}\mu ^2\frac d{d\mu ^2}Z_\Gamma=
\sum_{i=0}^\infty\gamma_\Gamma^{(i)}\left( \frac{\alpha_s}{4\pi }\right)^{i+1}\; ,
\ee 
where $\alpha_s=g^2/4\pi$ is the strong coupling constant.
If a renormalization scheme
which preserves the chiral Ward Identities (WI) is chosen 
(such as $\MSbar$ or RI), then
\begin{eqnarray}
& & \gamma_A =  \gamma_V = 0\; ,\\
& & \gamma_P =  \gamma_S = - \gamma_m\; ,\nonumber   
\label{eq:AWIg}
\end{eqnarray}
where $\gamma_m$ is the anomalous dimension of the quark mass.
Moreover the $\gamma_\Gamma^{(0)}$s are renormalization-scheme independent 
and gauge invariant and their values are reported in Table~\ref{tab:bil}.\\
\indent The anomalous dimensions are properties of the 
renormalized operators: they are independent of the 
regularization scheme adopted and depend only on 
the renormalization conditions imposed (the renormalization scheme) and 
the renormalized coupling constant used. For a given operator, defining the 
renormalization conditions is equivalent to fixing $\gamma_\Gamma(\alpha_s)$, i.e.
$\gamma_\Gamma(\alpha_s)$  uniquely determines the scheme in a regularization invariant way. 
Therefore, $\gamma_\Gamma$ can be computed in a simpler regularization, like Naive 
Dimensional Regularization (NDR), provided that the renormalization
conditions are the same. The evolution of the renormalized operators is determined by solving
the Renormalization Group Equations and, in a given scheme, is 
\be
\widehat O_\Gamma(\mu) = \frac{c_\Gamma(\mu)}{c_\Gamma(\mu_0)}O_\Gamma(\mu_0)   \; ,
\ee
where $c_\Gamma(\mu)$ at Next-to-Leading Order (NLO) 
\be
c_\Gamma(\mu)  = \left(\alpha_s(\mu) \right)^{\gamma_\Gamma^{(0)}/(2\beta_0)}
\left[ 1 +\frac{\alpha_s(\mu)}{4\pi} 
\left( \frac{\gamma^{(1)}_\Gamma}{2\beta_0} -
\frac{\beta_1\gamma^{(0)}_\Gamma}{2\beta_0^2} \right) \right] 
\label{eq:calfa}
\ee
is scheme dependent. In 
eq.~(\ref{eq:calfa}) $\beta_0$ and $\beta_1$ are the first two coefficients of the 
$\beta$ function of QCD. For $\gamma_\Gamma^{(1)}$ in RI or $\MSbar$
see for example \cite{bilrgi}. The Renormalization Group 
Invariant (RGI) operators can be defined as \cite{bilrgi}-\cite{lurgi}
\be
\widehat O_\Gamma^{\mbox{\scriptsize RGI}} = \frac{1}{c_\Gamma(\mu)} \widehat
O_\Gamma(\mu) = Z^{\mbox{\scriptsize RGI}}_\Gamma(a) O_\Gamma(a)\; ,
\ee
where 
\be
Z^{\mbox{\scriptsize RGI}}_\Gamma(a) = \frac{Z_\Gamma(\mu
  a)}{c_\Gamma(\mu)}\; .  
\ee
$\widehat O_\Gamma^{\mbox{\scriptsize RGI}}$ is 
independent of the scheme, scale and gauge chosen to renormalize 
the operators, while $Z^{\mbox{\scriptsize RGI}}_\Gamma(a)$ depends only 
on the regularization and not on the scheme, scale and 
gauge. 

As discussed in the introduction, 
it is very helpful to separate computations 
which can be performed using only lattice perturbation theory 
from those done in the continuum. In this regard the 
RI renormalization scheme proposed in \cite{RI,NP} is the optimum choice:
it preserves all the relevant symmetries (chirality and switch, see below) and 
it can also be easily implemented non-perturbatively \cite{NP}.
The matching with a ``continuum'' renormalization scheme, for example 
$\MSbar$, remains necessary 
because almost all the Wilson coefficients are computed 
in the continuum, but
it can be done using continuum perturbation theory only.
Therefore lattice and continuum regularizations  
are used independently taking full advantage 
of their properties. In the following we will indicate the 
renormalization scheme adopted with a superscript in the 
renormalization quantities, i.e. $Z_\Gamma^{\RI}$ and $Z_\Gamma^{\msbar}$  
for RI and $\MSbar$ schemes respectively. 

Using the Feynman rules defined in Appendix \ref{appb}, we computed the 
self energy of the quark propagator and the amputated Green's functions of bilinear operators 
between off-shell quark states at one loop 
in perturbation theory in a generic covariant gauge. The full 
expressions we obtained are reported in Appendix \ref{appd} and they 
can be used to impose any renormalization condition in any
covariant gauge.
\begin{table}
\begin{center}
\begin{tabular}{||l|l|rrrr||}
\hline\hline
\multicolumn{2}{||c|}{   }&$\psi$ & S  & V & T \\
\hline\hline
\multicolumn{6}{||c||}{Continuum PT} \\
\hline
\multicolumn{2}{||c|}{$\bar\gamma^{(0)}$}      & 0 & -6 & 0 & 2 \\
\multicolumn{2}{||c|}{$\Delta^{\msbar,\ri}$}   & 0 & -4 & 0 & 0 \\
\hline\hline
 & $\rho$ &\multicolumn{4}{|c||}{Lattice PT} \\
\hline
                &     0.2 &-235.80762 & 1.31942 & 1.52122 &  1.58848 \\ 
                &     0.3 &-150.61868 & 1.89625 & 1.52277 &  1.39828 \\ 
                &     0.4 &-108.19798 & 2.38060 & 1.52448 &  1.23911 \\ 
                &     0.5 & -82.86081 & 2.80522 & 1.52637 &  1.10009 \\ 
                &     0.6 & -66.05227 & 3.18782 & 1.52845 &  0.97532 \\ 
                &     0.7 & -54.10921 & 3.53927 & 1.53074 &  0.86124 \\ 
                &     0.8 & -45.20179 & 3.86686 & 1.53329 &  0.75543 \\ 
                &     0.9 & -38.31447 & 4.17577 & 1.53611 &  0.65622 \\ 
$B$             &     1.0 & -32.83862 & 4.46989 & 1.53924 &  0.56236 \\ 
                &     1.1 & -28.38734 & 4.75224 & 1.54274 &  0.47290 \\ 
                &     1.2 & -24.70304 & 5.02527 & 1.54665 &  0.38711 \\ 
                &     1.3 & -21.60760 & 5.29104 & 1.55105 &  0.30438 \\ 
                &     1.4 & -18.97397 & 5.55135 & 1.55601 &  0.22422 \\ 
                &     1.5 & -16.70910 & 5.80783 & 1.56163 &  0.14623 \\ 
                &     1.6 & -14.74330 & 6.06201 & 1.56804 &  0.07005 \\ 
                &     1.7 & -13.02336 & 6.31544 & 1.57541 & -0.00460 \\ 
                &     1.8 & -11.50798 & 6.56970 & 1.58394 & -0.07798 \\ 
\hline\hline
\end{tabular}
\caption{Perturbative values in the continuum and on the lattice in the 
Landau gauge, where $\bar\gamma^{(0)}=\gamma^{(0)}/C_F$ while the $B$s 
contain the proper vertex contribution only.}
\label{tab:bil}
\end{center}
\end{table}
The RI scheme is imposed by taking the trace of the amputated 
Green's functions in the Landau gauge ($\alpha=0$) \cite{NP}. 
The renormalization constant of the quark field is then  
defined as
\be\label{eq:fieldRC}
Z^{\RI}_\psi= 
-i\frac{1}{48}\Tr \gamma_\mu\frac{\partial S^{-1}(p)}{\partial p_\mu}
\Bigg \vert _{p^2=\mu^2} \; ,
\ee
while the renormalization constants of the bilinear operators are given by
\be
(Z^{\RI}_\psi(\mu a))^{-1} Z^{\RI}_\Gamma(\mu a) 
\Tr\Big[P_\Gamma \Lambda_\Gamma (p,a)\Big]\Bigg\vert_{p^2=\mu^2}  = 1\; ,  
\label{eq:blRC}   
\ee
where the trace is over both color and spin indices and $P_\Gamma$ are 
suitable projectors on the tree-level operators defined in Appendix
\ref{appc}. In general, these renormalization conditions differ from the standard
$\RIb$ perturbative prescriptions (see for example \cite{guido}), 
where the renormalization constants are extracted only from terms
proportional to the tree-level matrix elements of the operators under
consideration. The differences are evident from the expressions 
of the quark propagator and the amputated Green's functions reported 
in Appendix \ref{appd}. In the standard procedure, finite terms such 
as $\pslash p_\mu/p^2$ in the vector current
are considered as part of the matrix elements of the operators, whereas
with the projectors they give additional finite contributions 
to the renormalization 
constants\footnote{Note that in the Landau gauge these terms are absent
both in the bilinears and in the four-fermion operators.}. 
These terms are the same on the lattice and in the continuum
and they cancel when one computes the difference between the continuum 
and the lattice renormalization constants, as in the standard perturbative 
procedure.\\
\indent Using eqs.~(\ref{eq:fieldRC}) and (\ref{eq:blRC}) and the results reported in 
Appendix \ref{appd} we obtained
\ba\label{eq:Zbil}
Z^{\ri}_\psi   & = &  1 - \frac{g_0^2}{16\pi^2} C_F
\left[\frac{\bar\gamma_\psi^{(0)}}{2}\log(\mu a)^2 
+ B_\psi \right]\nonumber\\
Z^{\ri}_\Gamma & = & 1 - \frac{g_0^2}{16\pi^2} C_F
\left[\frac{\bar\gamma_\Gamma^{(0)}}{2}\log(\mu a)^2 
+ B_\Gamma + B_\psi \right] \; ,
\ea
where the $\bar\gamma_\Gamma^{(0)}$ and $B$s are reported in
Table~\ref{tab:bil}. \\
\indent Once the operators have been renormalized in RI,
the matching with another given scheme $\chi$ is obtained with
\be
Z^{\chi}_\Gamma(\mu a) = c^{\chi}_\Gamma(\mu) Z^{\mbox{\scriptsize RGI}}_\Gamma(a) = 
\frac{c^{\chi}_\Gamma(\mu)}{c^{\ri}_\Gamma(\mu)} Z^{\ri}_\Gamma(\mu a)\; ,  
\ee    
where the $c^{\ri}_\Gamma(\mu)$ are reported in \cite{bilrgi} for 
the bilinears, defined with the same conventions adopted in this 
paper.\\
\indent To compare our results with \cite{vicari}, we report
explicitly the matching coefficients for the $\MSbar$ scheme.
The chiral WI imply
$Z_V^{\ri} = Z_V^{\msbar}$,
$Z_A^{\ri} = Z_A^{\msbar}$ and  
\begin{equation}
\frac{Z_S^{\ri}}{Z_P^{\ri}} = \frac{Z_S^{\msbar}}{Z_P^{\msbar}} = 1\; .
\label{eq:WIrel} 
\end{equation}  
A straightforward computation in NDR 
gives
\begin{equation}
\label{eq:matching}
Z_{\Gamma}^{\msbar}(a \mu)= \left[1-
C_F \dfrac{\alpha_s}{4\pi} (\Delta^{\msbar,\RI}_\Gamma + 
\Delta^{\msbar,\RI}_\psi )\right] Z_\Gamma^{\RI}(a \mu), 
\label{eq:Z_MSbar}
\end{equation} 
where the $\Delta^{\msbar,\RI}$s are reported in Table~\ref{tab:bil}.
In the $\MSbar$ scheme, the renormalization constants of gauge-invariant 
operators are independent of the gauge. Therefore, 
the dependence on the gauge and 
external states of the RI scheme cancels the corresponding dependence of 
the matching coefficients. Our results in $\MSbar$ are in 
agreement with those of Ref.~\cite{vicari}.

\section{Four-fermion operators}
\label{sec:4-fermions} 
In this section we introduce the most general set of dimension-six
four-fermion operators we are interested in, we analyze their mixing 
pattern exploiting only the symmetries of the underlying theory and 
finally we compute the necessary renormalization constants at 1-loop 
in perturbation theory. A very exhaustive non-perturbative analysis 
for Wilson fermions has been done in Ref.~\cite{tassos}. We will proceed on 
the same lines with the advantage of having the additional chiral symmetry 
in eq.~(\ref{eq:luscher}) which forbids mixing among operators which belong 
to multiplets with different chirality \cite{hasenfratz2}.\\
\indent The generic set of four-fermion operators we are interested in is
\ba\label{eq:basstart}
O_{\Gamma_A \Gamma_B} & = & 
\bar \psi_1\Gamma^{l}_A \psi_2 \cdot \bar \psi_3\Gamma^{l}_B \psi_4\\
O^F_{\Gamma_A \Gamma_B} & = & 
\bar \psi_1\Gamma^{l}_A \psi_4 \cdot \bar \psi_3\Gamma^{l}_B \psi_2\; ,\nonumber
\ea
where the flavors $\psi_1-\psi_4$ are all different and the  
$\Gamma^{l}_A$ are generic Dirac matrices with
$l$ representing the contracted Lorenz indices (if any).
The 20 operators in (\ref{eq:basstart}) form a complete 
basis of four-fermion operators.
Operators with the color indices contracted in a different way can 
be expressed as linear combinations of the ones in 
eq.~(\ref{eq:basstart}) using the color Fierz identity (\ref{eq:fiertzcol}) 
in Appendix \ref{appa}.\\
\indent The $O(a)$ discretization errors in on-shell four-fermion 
matrix elements can be removed by using the improved operator
\begin{equation}
\bar{\psi} \Big(1-\frac{1}{2\rho} a D_N\Big) \, \Gamma_A \, 
\Big(1-\frac{1}{2\rho} a D_N\Big) \psi \cdot  
\bar{\psi} \Big(1-\frac{1}{2\rho} a D_N\Big) \, \Gamma_B \, 
\Big(1-\frac{1}{2\rho} a D_N\Big) \psi .
\label{eq:improvedneuberger}
\end{equation}
This can be proven along the lines used in Refs.~\cite{qcdsf} for the case of 
two-quark operators. Here we just sketch the argument, which is valid 
for any reasonable action which satisfies the GWR. Starting from the 
Neuberger operator $D_N$, one can define the associated operator~\cite{chiu}
\begin{equation}
K_N = \Big(1-\frac{1}{2\rho} a D_N\Big)^{-1} D_N ,
\label{eq:transfkd}
\end{equation}
which has the same chiral properties as the continuum Dirac operator, i.e.
$\{ K_N, \gamma_5 \}=0$. The propagator $K^{-1}_N$ is 
free of $O(a)$ corrections and, although otherwise not well-behaved 
and so not useful in practice, it turns out to be very useful 
for the construction of operators that are improved. In fact, 
the four-fermion correlation function   
 \ba
& & \Big\langle \frac{1}{K_N} \Gamma_A \frac{1}{K_N} \cdot 
\frac{1}{K_N} \Gamma_B \frac{1}{K_N} \Big\rangle  = \\
& & \Big\langle \frac{1}{D_N} \Big(1-\frac{1}{2\rho} a D_N\Big) \Gamma_A 
\Big(1-\frac{1}{2\rho} a D_N\Big) \frac{1}{D_N} \cdot
\frac{1}{D_N} \Big(1-\frac{1}{2\rho} a D_N\Big) \Gamma_B
\Big(1-\frac{1}{2\rho} a D_N\Big) \frac{1}{D_N} \Big\rangle\;  \nonumber
\ea
is automatically improved. In the last line we have used 
eq.~(\ref{eq:transfkd}) to re-express the Green's function in terms of 
the Neuberger propagator, and from this we can read off the improved 
operator.
Thus, the Ginsparg-Wilson relation highly simplifies the improvement
of the four-fermion operators at all orders in perturbation theory. 
Moreover, the renormalization factors for corresponding improved and 
unimproved operators are the same. This happens because in 1-loop amplitudes
a $D_N$ factor can combine with a quark propagator but,  
since it has an $a$ in front and (contrary to the Wilson case) there is no 
$1/a$ factor in the propagator, as additive mass renormalization is forbidden 
by chiral symmetry, its contribution to the renormalization factors 
is zero~\cite{vicari}. 

The symmetries relevant for studying the mixing of the operators that we consider are  
Parity (${\cal P}$),
the Switch (${\cal S}$) transformation $\psi_2\leftrightarrow \psi_4$
and chiral symmetry in eq.~(\ref{eq:luscher}). They 
allow splitting the original basis into smaller independent multiplets. 
For parity-conserving (${\cal P}=1$) operators it is useful to define 
the following bases 
\ba\label{eq:4fermdefinitions}
Q^{\pm}_1 & = & O^\pm_{VV} +  O^\pm_{AA}\nonumber\\
Q^{\pm}_2 & = & O^\pm_{VV} -  O^\pm_{AA}\nonumber\\
Q^{\pm}_3 & = & O^\pm_{SS} -  O^\pm_{PP}\nonumber\\
Q^{\pm}_4 & = & O^\pm_{SS} +  O^\pm_{PP}\nonumber\\
Q^{\pm}_5 & = & O^\pm_{TT} \; ,
\ea
and for the parity-violating (${\cal P}=-1$) 
\ba\label{eq:4fermdefinitions2}
{\cal Q}^{\pm}_1 & = & O^\pm_{VA} +  O^\pm_{AV}\nonumber\\
{\cal Q}^{\pm}_2 & = & O^\pm_{VA} -  O^\pm_{AV}\nonumber\\
{\cal Q}^{\pm}_3 & = & - O^\pm_{SP} +  O^\pm_{PS}\nonumber\\
{\cal Q}^{\pm}_4 & = & O^\pm_{SP} +  O^\pm_{PS}\nonumber\\
{\cal Q}^{\pm}_5 & = & O^\pm_{T\widetilde T} \; ,
\ea
where
\ba
O^{\pm}_{\Gamma_A \Gamma_B} & = & \frac{1}{2} \left [
O_{\Gamma_A \Gamma_B} \pm O^F_{\Gamma_A \Gamma_B} \right ] \; ,
\label{eq:qgamgam}
\ea
and $\widetilde T = \sigma_{\mu\nu} \gamma_5$. Since the lattice preserves
parity, the two sets of operators
(\ref{eq:4fermdefinitions}) and (\ref{eq:4fermdefinitions2}) renormalize
independently. Using continuous chiral 
transformations it is easy to show that,
at variance with Wilson fermions, they renormalize with the same 
renormalization matrices \cite{tassos}. We have explicitly 
verified this property at one loop in perturbation theory. 
In the following we will then consider only the parity-conserving 
operators in 
eqs.~(\ref{eq:4fermdefinitions}). 
Among them, the five $Q^{+}_i$ are left invariant under a switch transformation (${\cal S}=1$), 
while the $Q^{-}_i$ change sign (${\cal S}=-1$). Therefore the two sets 
renormalize independently as
\be   
\widehat Q^{\pm}_i(\mu) =    
\widehat Z^{\pm}_{ij}(\mu a ) Q^{\pm}_j(a)\; ,   
\label{eq:zodef2}   
\ee 
where $\widehat Z^{\pm} $ are the renormalization constant matrices.\\
\indent The operators of the bases (\ref{eq:4fermdefinitions}) do not belong,
in general, to irreducible representations of the chiral group. Nevertheless
chiral symmetry imposes many constraints on the matrix of the
renormalization constants when a scheme which preserves 
it together with the switch symmetry is chosen.
The most general mixing matrix under these constraints reads
\be\label{eq:zete}
\widehat Z^{\pm} = \pmatrix{ 
Z^{\pm}_{11}&      0     &      0         &      0      & 0\cr
     0      & Z_{22}     & \pm Z_{23}     &      0      & 0\cr
     0      & \pm Z_{32} &    Z_{33}      &      0      & 0\cr
     0      &      0     &      0         & Z^{\pm}_{44}& Z^{\pm}_{45}\cr
     0      &      0     &      0         & Z^{\pm}_{54}& Z^{\pm}_{55}\cr} \; . 
\ee
Note that for $i,j=2,3$, $Z^+_{ij}$ and $Z^-_{ij}$
are related. Therefore for Neuberger fermions there are 
only 14 independent renormalization constants, which would instead become 64 if the 
Wilson action were used\footnote{In the Wilson case the renormalization matrices of 
parity-conserving and parity-violating operators are different \cite{tassos}. 
Moreover all the entries of the analogous of the matrices in eq.~(\ref{eq:zete}) are 
independent and non-zero. This results in $25\times2 + 14=64$ independent
constants. It is interesting to note that using suitable Ward
Identities, some parity-conserving matrix elements can be related to 
the parity-violating ones \cite{martibello}.}. 
The particular structure of the matrix in 
eq.~(\ref{eq:zete}) also simplifies the implementation of non-perturbative 
techniques \cite{NP,luscher_np}.

Analogously to the bilinears, denoting by  $x_1,x_3$ and $x_2,x_4$  
the coordinates of the  
outgoing and incoming quarks respectively, the   
four-point Green's functions are defined as
\begin{equation}
G^{\pm}_i(x_1,x_2,x_3,x_4)=
\<\psi_1(x_1)\bar\psi_2(x_2) Q^{\pm}_i(0)
\psi_3(x_3)\bar\psi_4(x_4)\>\, ,  
\label{eq:G_Gamma(x)}
\end{equation}
where $\<\cdots\>$ denotes the vacuum expectation value.
Note that $G^{\pm}_i$ depends implicitly on the four color
and Dirac indices carried by the external fermion fields. 
The Fourier transform of the Green's function 
(\ref{eq:G_Gamma(x)}) at equal external momenta $p$ is defined as
\begin{equation}\label{eq:G_Gamma(p)}
G^{\pm}_i(p)
= \int dx_1 dx_2 dx_3 dx_4 
e^{-ip(x_1 + x_3 - x_2 - x_4)} G^{\pm}_{i}
(x_1,x_2,x_3,x_4)\; .
\end{equation}
The corresponding amputated correlation functions is defined as 
\be
\Lambda^{\pm}_i(p) = S^{-1}(p) S^{-1}(p) G^{\pm}_i(p)  S^{-1}(p) S^{-1}(p)\; .
\ee
From the above definitions we find for the renormalized Green's functions   
\ba   
\widehat G^{\pm}_{i}(p) & = & Z^2_\psi \widehat Z^{\pm}_{ij} G^{\pm}_{j}(p)    
\nonumber \; ,\\  
\widehat \Lambda^{\pm}_{i}(p) & = & Z^{-2}_\psi \widehat Z^{\pm}_{ij} \Lambda^{\pm}_{j}(p) \; .   
\label{eq:cl}   
\ea
The anomalous dimension matrices
are defined as 
\be
\widehat \gamma^{\pm}(\alpha_s)\equiv -2 (\widehat Z^{\pm})^{-1}\mu^2
\frac{d}{d\mu^2}\widehat Z^{\pm} = 
\sum_{i=0}^\infty\widehat\gamma^{\pm(i)}\left( \frac{\alpha_s}{4\pi }\right)^{i+1}\; 
\ee
and, if the renormalization scheme preserves chiral 
symmetry, they have the same structure 
as in eq.~(\ref{eq:zete}). At first order in $\alpha_s$ we obtain 
\ba
\gamma^{+(0)}_{11} = -\frac{2}{3}\bar\gamma^{(0)}_S  
& & \gamma^{-(0)}_{11} = \frac{4}{3}\bar\gamma^{(0)}_S  \\  
\gamma^{(0)}_{22} = -\frac{1}{3}\bar\gamma^{(0)}_S  
& & \gamma^{(0)}_{23} =   - 2 \bar\gamma^{(0)}_S \nonumber\\
\gamma^{(0)}_{32} =  0 
& & \gamma^{(0)}_{33} =   \frac{8}{3}\bar\gamma^{(0)}_S \nonumber\\
\gamma^{+(0)}_{44} =  \frac{23 \bar\gamma^{(0)}_S  + 9 \bar\gamma^{(0)}_T}{12}
& & \gamma^{-(0)}_{44} =\frac{41 \bar\gamma^{(0)}_S - 9 \bar\gamma^{(0)}_T}{12}\nonumber\\
\gamma^{+(0)}_{45} =  \frac{ - \bar\gamma^{(0)}_S  + \bar\gamma^{(0)}_T}{12} 
& & \gamma^{-(0)}_{45} =   \frac{5(\bar\gamma^{(0)}_S  - \bar\gamma^{(0)}_T)}
{12}\nonumber\\
\gamma^{+(0)}_{54} = \frac{5(\bar\gamma^{(0)}_S  - \bar\gamma^{(0)}_T)}{4}
& & \gamma^{-(0)}_{54} = \frac{ - \bar\gamma^{(0)}_S  + \bar\gamma^{(0)}_T}{4} \nonumber\\
\gamma^{+(0)}_{55} =  \frac{-9\bar\gamma^{(0)}_S  + 41 \bar\gamma^{(0)}_T}{12} 
& & \gamma^{-(0)}_{55} =   \frac{9\bar\gamma^{(0)}_S  + 23
  \bar\gamma^{(0)}_T}{12}\; , \nonumber  
\ea
which agree with \cite{ciuco,misiak}, where the anomalous dimension matrices at two 
loops in the same bases can also be found\footnote{The operators $Q^{\pm}_5$ used in this 
paper and defined in equation (\ref{eq:4fermdefinitions}) correspond to $1/4$ of those
defined in \cite{ciuco,misiak}.}.\\
\indent The renormalization group evolution
of the four-fermion operators, in a scheme which preserves chiral and the 
switch symmetry, is 
\be
\widehat Q^{\pm}(\mu) = \widehat{w}^{\pm}(\mu) [\widehat{w}^{\pm}(\mu_0)]^{-1}  
 \widehat Q^{\pm}(\mu_0)\; , 
\ee  
where in this case $\widehat{w}^{\pm}(\mu)$ are matrices which depend only 
on the anomalous dimensions and $\alpha_s$. Their expression at NLO
in a generic scheme can be found in \cite{ciuco,misiak,NoiDELTAS=2}.
Analogously to the bilinears, the RGI operators can be defined 
as \cite{NoiDELTAS=2}
\be
[\widehat Q^{\pm}]^{\mbox{\scriptsize RGI}} = [\widehat{w}^{\pm}(\mu)]^{-1}
\widehat Q^{\pm}(\mu) = [\widehat Z^{\pm}(a)]^{\mbox{\scriptsize RGI}} Q^{\pm}(a)\; ,
\ee
where 
\be
[\widehat Z^{\pm}(a)]^{\mbox{\scriptsize RGI}} = 
[\widehat{w}^{\pm}(\mu)]^{-1} \widehat Z^{\pm}(\mu a)\; .
\ee
As for the scheme, scale and gauge dependences, these definitions have the
same properties as for bilinears.
\begin{figure}[htb]
\begin{center}
\caption{\it{An example of Fierz and charge conjugation rearrangements for 
Feynman graphs of four-fermion operators.}}
\label{fig:feynman}
\vskip 0.3cm
\includegraphics[height=4cm,width=12cm]{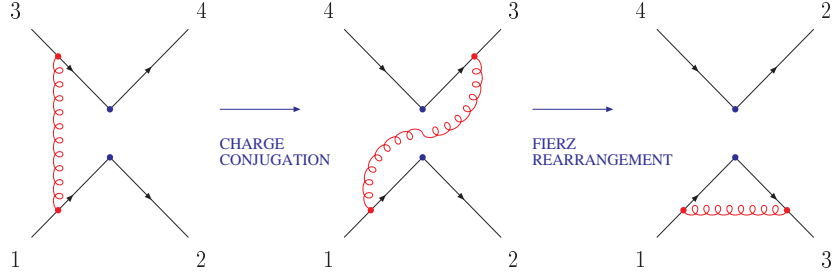}
\end{center}
\end{figure}

We have computed the amputated Green's functions of the 
four-fermion operators  at one loop in two different ways. 
In the first one we use the Fierz and charge conjugation rearrangements 
to connect the proper vertices of the four-fermion 
operators to the bilinear ones. In the second, the calculation
has been performed computing the four-quark diagrams directly without any 
rearrangement of quark legs. To carry out the analytic 
calculations we have used a set of routines written in the symbolic 
manipulation language FORM, which are a generalization to the overlap case 
of the ones used in Refs.~\cite{stefano}. Many of the features of those 
calculations are also present here. We have numerically integrated the 
analytic FORM outputs for some values of the $\rho$ parameter and we have 
successfully checked the results obtained with the first method. This also turns
out to be a non-trivial check of the numerical integrations 
relevant in the Neuberger regularization.
The full expressions for the one-loop amputated Green's functions that  
we have obtained in a generic covariant gauge are 
reported in Appendix \ref{appd}, and they can be used to impose
any set of renormalization conditions.
  
Once the results have been rotated in the basis (\ref{eq:4fermdefinitions}), 
the RI renormalization conditions are imposed by taking the trace the amputated 
Green's functions in the Landau gauge \cite{tassos}
\be
\Tr \Pj^{\pm}_{i} \widehat\Lambda_k^{\pm} = \delta_{ik}\qquad (i,k=1,\cdots,5)\; ,
\ee 
where the projectors $\Pj^{\pm}_{i}$ are defined in Appendix \ref{appc}.
Using the wave function renormalization $Z^{\RI}_\psi$ defined in 
eq.~(\ref{eq:fieldRC}) and the definition in (\ref{eq:cl}) we obtain 
\be
[\widehat Z^{\pm}]^{\ri} =  1 - \frac{g_0^2}{16\pi^2}
\left[\frac{\widehat \gamma^{(0)}}{2}\log(\mu a)^2 
+ \widehat B^{\pm} + 2 C_F B_\psi \right] \; ,\nonumber\\
\ee
where
\be
\widehat B^{\pm} = \pmatrix{
B^{\pm}_{11} &      0      &      0      &      0      & 0\cr
     0       &  B_{22}     & \pm B_{23}  &      0      & 0\cr
     0       & \pm B_{32}  &    B_{33}   &      0      & 0\cr
     0       &      0      &      0      & B^{\pm}_{44} & B^{\pm}_{45}\cr
     0       &      0      &      0      & B^{\pm}_{54} & B^{\pm}_{55}\cr} \; , 
\ee
and the $B^{\pm}_{ij}$ are 
\ba\label{eq:Bs}
B^{+}_{11} = \frac{-2(B_S - 5 B_V)}{3}
& & B^{-}_{11} = \frac{4(B_S + B_V)}{3}\\  
B_{22} = -\frac{B_S}{3} + 3 B_V  
& & B_{23} = -2 (B_S - B_V)  \nonumber\\
B_{32} =   0 
& & B_{33} =   \frac{8 B_S}{3} \nonumber\\
B^{+}_{44} =  \frac{ 23 B_S + 9 B_T}{12} 
& & B^{-}_{44} =  \frac{ 41B_S  - 9 B_T}{12}     \nonumber\\
B^{+}_{45} = \frac{ - B_S + B_T}{12}   
& & B^{-}_{45} =   \frac{ 5 (B_S  - B_T)}{12}   \nonumber\\
B^{+}_{54} = \frac{ 5 (B_S  - B_T)}{4}   
& & B^{-}_{54} =   \frac{ - B_S + B_T}{4}   \nonumber\\
B^{+}_{55} = \frac{ -9 B_S  + 41 B_T}{12}     
& & B^{-}_{55} =  \frac{ 9 B_S + 23 B_T}{12}\; ,\nonumber  
\ea
where the $B_\Gamma$'s are reported in Table~\ref{tab:bil}.
The relations above are valid in the Landau gauge for any action which 
preserves chiral symmetry, where $B_A=B_V$ and $B_P=B_S$. 

The Wilson coefficients and the four-fermion operators 
of the weak effective Hamiltonian are often defined in one of 
the $\MSbar$ schemes.
Unfortunately the definition of the $\MSbar$ scheme for the 
four-fermion operators  is not unique even if we consider the 
Naive Dimensional Regularization only. This is a source of confusion in 
the literature, where quite often one encounters comparisons of matrix 
elements that have been computed in different schemes. 
Incidentally, we note that the 
$\MSbar$-NDR scheme used in the lattice calculation of Ref.~\cite{sharpe} 
differs from the one used in ~\cite{ciuco}, which moreover is not the one 
adopted in \cite{misiak}.  
In some cases, differences between various schemes may be 
numerically large. On the other hand the matrix elements of the 
four-fermion operators in a given scheme are useful 
only if they are matched with the Wilson coefficients computed in the same 
scheme. From the perturbative computation of the Wilson coefficients 
in a given scheme $\chi$, it is straightforward to extract 
the evolution operator $[\widehat{w}^{\pm}(\mu)]^{\chi}$ at the NLO. 
Then the corresponding renormalization constants are given by
\be
[\widehat Z^{\pm}(\mu a)]^{\chi} = [\widehat{w}^{\pm}(\mu)]^{\chi} 
[\widehat Z^{\pm}(a)]^{\mbox{\scriptsize RGI}} = 
[\widehat{w}^{\pm}(\mu)]^{\chi} [\widehat{w}^{\pm\; \RI}(\mu)]^{-1}
[\widehat Z^{\pm}(\mu a)]^{\RI}\; ,
\ee
where $[\widehat{w}^{\pm}(\mu)]^{\RI}$ can be found in 
\cite{ciuco,NoiDELTAS=2,misiak}. This is equivalent to
using only RGI Wilson coefficients and operators as proposed in \cite{NoiDELTAS=2}.    

\section{Conclusions}\label{sec:conc}
The matrix elements of dimension-six four-fermion operators without power
subtractions are the primary ingredients for studying many interesting 
phenomenological problems in weak interactions, among which the most important are the 
predictions of the $K^0-\bar{K}^0$ and $B^0-\bar{B}^0$ mixings in the
Standard Model and beyond. The renormalization 
constants of these operators are also relevant for studying the 
$\Delta I =1/2$ rule and computing the CP-violation 
parameter $\epsilon^\prime/\epsilon$.\\ 
\indent In this paper we have studied in detail the mixing of these 
operators using the overlap lattice regularization 
proposed by Neuberger and computed their renormalization factors at one loop
in perturbation theory. The computations were 
done in two independent ways: in one case, using Fierz and charge conjugation 
rearrangements, the four-fermion Green's functions are given 
in terms of the one loop results for the bilinear 
operators; in the other
case the Feynman diagrams are computed directly using FORM codes.\\
\indent We have shown that operators belonging to 
different chiral representations do not mix among themselves.
We have explicitly verified that, as expected, the 
mixing matrices for the parity-violating and parity-conserving 
operators are the same. Furthermore, the improvement of the matrix 
elements is highly simplified by the Ginsparg-Wilson relation and the renormalization constants are 
the same as the unimproved case. In the Wilson formulation, whether improved or not, 
the construction of renormalized operators requires subtractions of operators with wrong
chirality. These become very severe in the case of power divergences, 
i.e. $\Delta I=1/2$, $\epsilon'/\epsilon$, etc. We believe that 
the overlap regularization represents a very promising approach for solving these 
long-standing problems. 

\section*{Acknowledgment}
We thank  G.~Martinelli for a very illuminating discussion on the 
chiral properties of the operators. L. G. thanks also C.~Hoelbling,  V.~Lubicz, C.~Rebbi
and M.~Schwetz for stimulating discussions. S. C. has been supported 
in part by the U.S. Department of Energy (DOE) under cooperative 
research agreement DE-FC02-94ER40818. L. G. has been supported in part under 
DOE grant DE-FG02-91ER40676. 

\begin{appendix}
\section{Feynman Rules}\label{appb}
In this appendix we report the Feynman rules used in the computations.
Some abbreviations of functions occurring on the lattice are useful:
\ba
\hatk   & = & \frac{2}{a}\sin(\frac{k_\mu a}{2}) \qquad
\hatkk    = \sum_{0}^3 \hat{k}_\mu^2\nonumber\\
\bark   & = & \frac{1}{a} \sin(k_\mu a) \qquad
\barkk    = \sum_{0}^3 \bar{k}_\mu^2 \; . 
\ea
The gluonic propagator in a generic covariant gauge is defined as
\be
G_{\mu\nu}(k) = \left(\delta_{\mu\nu} 
+ (\alpha - 1 ) \hatk\hat{k}_\nu/\hatkk \right)/\hatkk \; ,
\ee
where $\alpha$ is the gauge parameter.

Let us define some useful matrices in Dirac space:
\ba
X_0(p) & = & i\gamma_\mu \bar p_\mu -\frac{\rho}{a} + \frac{ar}{2}\hat{p}^2\\
X_{1\mu}(p+p') & = & -ig_0\left[\gamma_\mu \cos\Big(\frac{p_\mu + p'_\mu}{2}a\Big) 
-ir\sin\Big(\frac{p_\mu + p'_\mu}{2}a\Big)\right]\nonumber\\
X_{2\mu\nu}(p+p') & = & -a \frac{g_0^2}{2}\delta_{\mu\nu}\left[r\cos\Big(\frac{p_\mu +
    p'_\mu}{2}a\Big) - i\gamma_\mu \sin\Big(\frac{p_\mu + p'_\mu}{2}a\Big)\right]\; .\nonumber 
\ea
The fermionic propagator of the overlap action is 
\be
S^{(0)}_N(p) = \frac{a}{2\rho}\left[\frac{-i\gamma_\rho\bar p_\rho}{\omega(p)+b(p)} 
             + 1 \right] =
\frac{a}{2\rho}\frac{X_0^\dagger(p) + \omega(p)}{\omega(p) + b(p)}\; ,\label{eq:D_0^{-1}}
\ee
with
\ba
b(p) &=& \frac{ar}{2}\hatpp - \frac{\rho}{a}\; ,\\
  \omega(p) &=& \sqrt{\barpp + 
\left(\frac{ra}{2}\hatpp -\frac{\rho}{a}\right)^2}\; .\label{eq:w}
\end{eqnarray}
For $ 0 < \rho < 2 r$ the propagator $S^{(0)}_N(p)$ exhibits a massless 
pole only when $p_\mu =0$, and there are no doublers. 

The vertices of the overlap Dirac operator are
\begin{eqnarray}
V^{(1)}_{\mu}(p,p') & = &  \frac{\rho}{\omega(p') +\omega(p)}\left\{ 
X_{1\mu}\left(p+p'\right) - 
\frac{X_0(p')}{\omega(p')}X_{1\mu}^\dagger\left(
p+p'\right)\frac{X_0(p)}{\omega(p)}\right\}
\end{eqnarray}
and 
\begin{eqnarray}
V^{(2)}_{\mu\nu}(p,p')  & = & \left[ \frac{\rho}{\omega(p') + \omega(p)}
\left\{ X_{2\mu\nu}\left(p+p'\right) - \frac{X_0(p')}{\omega(p')}X_{2\mu\nu}^\dagger\left(
p+p' \right)\frac{X_0(p)}{\omega(p)} \right\}\right.\nonumber\\ 
 & + & \frac{\rho}{2 \{\omega(p') + \omega(p)\}\{\omega(p) + 
\omega(p+k)\}\{\omega(p+k) + \omega(p')\}}\\
 & \times &  \left\{ X_0(p')X^\dagger_{1\mu}\left( 
p+p'+k\right)X_{1\nu}\left(2 p + k \right) + 
   X_{1\mu}\left(p+p'+k\right)X_0^\dagger(p+k) 
X_{1\nu}(2p + k)\right.\nonumber\\
 & + &   X_{1\mu}\left(p+p'+k \right)
  X_{1\nu}^\dagger\left(2p + k\right)X_0(p)\nonumber \\
& - &   \left.\left.\frac{\omega(p') +\omega(p) +\omega(p+k)}
{\omega(p')\omega(p)\omega(p+k)}X_0(p')X_{1\mu}^\dagger
\left(p+p'+k\right)X_0(p+k)
X_{1\nu}^\dagger\left(2p + k\right)X_0(p)\right\}\right]\; .\nonumber 
\end{eqnarray}    

\section{Notations and Fierz Transformations}\label{appa}
In this appendix we fix our notations for the color and spin matrices. We also 
report the formul\ae~used for Fierz transformations in color and Dirac 
space.\\
The Gell-Mann group generators of the $SU(N_c)$ Lie algebra  are denoted by 
$t^A,A=1,\dots,N^2_c-1$. They are Hermitian, traceless $N_c\times N_c$ matrices and 
are normalized according to
\be
\Tr(t^A t^B) = \frac{1}{2}\delta^{AB}\; .
\ee
They satisfy the commutation relations 
\be
[t^A,t^B]   =  i f^{ABC}t^C \; ,\label{eq:comm}\\
\ee
where summation over repeated indices is implied. The structure 
constants $f^{ABC}$ are completely antisymmetric and real.
With these conventions the completeness relation for
the $t$-generators reads
\be
\sum_{A=1}^{N_c^2-1} (t^A)_{ab} (t^A)_{cd} = \frac{1}{2} 
\left(\delta_{ad}\delta_{bc} - \frac{1}{N_c}\delta_{ab}\delta_{cd}
\right) \; ,
\label{eq:fiertzcol}
\ee
and using the above formulas we get 
\ba
\sum_{A=1}^{N_c^2-1} (t^A t^A)_{ab} & = & C_F\delta_{ab}\; ,\\  
\sum_{C,D=1}^{N_c^2-1}f^{ACD}f^{BCD} & = &C_A\delta^{AB}\; ,
\ea
with $C_F = (N_c^2-1)/2N_c$ and $C_A = N_c$.

The complete basis of 16 Euclidean Dirac $4\times 4$ 
matrices is denoted by
\begin{equation}
\Gamma
=\{\id,\gamma_{\mu}, \sigma_{\mu\nu}, \gamma_{\mu}\gamma_5,\gamma_5\}
\equiv \{S,V,T,A,P\},
\end{equation}
where $\gamma_i$ are the usual Euclidean Dirac matrices in four dimensions
and 
\be
\sigma_{\mu\nu}=\frac{1}{2}[\gamma_{\mu},\gamma_{\nu}]\; .
\ee
The charge conjugation matrix satisfies
\be
C \gamma^T_\mu C^{-1} = -\gamma_\mu\; ,
\ee
and in our basis $C=\gamma_0 \gamma_2$.

Repeated $\Gamma$ matrices imply summation of their Lorenz indices (if any); 
for example $VV = \sum_\mu \gamma_\mu \otimes \gamma_\mu$, 
$TT = \sum_{\mu<\nu}\sigma_{\mu \nu} \otimes \sigma_{\mu \nu}$, where
the sum is over the 6 independent $\sigma_{\mu \nu}$ matrices only. With these
conventions the $\Gamma$ matrices are normalized as
\begin{equation}
SS=1,\ VV=4,\ TT=-6,\  AA=-4,\ PP=1\; ,
\end{equation}
where summation over Dirac indices is understood. 

The Fierz transformation of the Dirac indices of a four-fermion operator is defined as 
\begin{equation}
{\Gamma} \otimes {\Gamma} \equiv \Gamma_{\alpha\beta} \otimes
\Gamma_{\gamma\delta} \rightarrow \left[ {\Gamma} \otimes
{\Gamma} \right]^{F_D} \equiv \Gamma_{\alpha\delta} \otimes \Gamma_{\gamma\beta}\; .
\end{equation}
The Euclidean Fierz-transformed Dirac tensor products
$\left[ {\Gamma} \otimes {\Gamma} \right]^{F_D}$ can be 
re-expressed as a linear combination of the complete set of the
original tensor products $\Gamma \otimes \Gamma$ as follows :
\begin{equation}
\left(\begin{array}{c} 
\left[ {{S}} \otimes {{S}} \right]^{F_D} \\
\left[ {{V}} \otimes {{V}} \right]^{F_D} \\
\left[ {{T}} \otimes {{T}} \right]^{F_D} \\
\left[ {{A}} \otimes {{A}} \right]^{F_D} \\
\left[ {{P}} \otimes {{P}} \right]^{F_D} \\
\end{array}\right)
=-\frac{1}{4}
\left(\begin{array}{rrrrr}
1 &  1 & -1 & -1 & 1  \\
4 & -2 &  0 & -2 & -4  \\
-6 &  0 & -2 & 0 & -6  \\
-4 & -2 & 0 & -2 & 4  \\
 1 & -1 & -1 & 1 & 1  
\end{array}\right)
\left(\begin{array}{c} 
{S} \otimes {S} \\
{V} \otimes {V} \\
{T} \otimes {T} \\
{A} \otimes {A} \\
{P} \otimes {P} \\
\end{array}\right)\; .
\label{eq:fidir}
\end{equation}
The overall minus sign is due to the anti-commutativity of the Fermi fields.

\section{Projectors for Bilinears and Four-Fermion Operators}\label{appc}
In this appendix we define the projectors used to impose the RI 
renormalization conditions for the bilinear and
four-fermion operators.

For bilinears the projectors are defined as 
\ba   
\Pj_S = \frac{1}{12} \, I \qquad  && \qquad   
\Pj_P = \frac{1}{12} \, \gamma_5 \nonumber \\   
\Pj_V = \frac{1}{48} \, \gamma_\mu \qquad && \qquad   
\Pj_A = \frac{1}{48} \, \gamma_5 \gamma_\mu \nonumber \\
 \Pj_T =  -\frac{1}{72} \, \sigma_{\mu\nu}\; , &&   
\label{eq:proj}   
\ea
where the identity color matrix is not shown.

If we define
\begin{equation}
\Pj_{\Gamma\Gamma}\equiv
\Gamma \otimes \Gamma \; ,
\end{equation}
its trace on a generic four-fermion amputated Green's function 
is defined as
\cite{tassos}
\begin{equation}
\Tr\Pj_{\Gamma\Gamma} 
\Lambda_{\Gamma\Gamma} (p) =
\Gamma_{\sigma\rho}\otimes \Gamma_{\sigma'\rho'}
 \Lambda_{\Gamma\Gamma}(p)^{RRR'R'}_{\rho\sigma\rho'\sigma'},
\label{eq:proj_def_1}
\end{equation}
where the color and spinor indices are explicitly reported and the 
trace is taken over spin and color. The projectors $\Pj^\pm_k$
for the parity conserving operators $Q^\pm_k$ are \cite{tassos}
\begin{eqnarray}
&& \Pj^{\pm}_1 \equiv + \frac{1}{64N_c(N_c\pm 1)}(\Pj_{VV}+\Pj_{AA}) \nn \\
&& \Pj^{\pm}_2 \equiv + \frac{1}{64(N_c^2-1)}(\Pj_{VV}-\Pj_{AA}) 
               \pm \frac{1}{32N_c(N_c^2-1)}(\Pj_{SS}-\Pj_{PP}) \nn \\
&& \Pj^{\pm}_3 \equiv \pm \frac{1}{32N_c(N_c^2-1)}(\Pj_{VV}-\Pj_{AA})
               +  \frac{1}{16(N_c^2-1)}(\Pj_{SS}-\Pj_{PP})  
\label{eq:projqpc} \\
&& \Pj^{\pm}_4 \equiv + \frac{(2N_c\pm 1)}{32N_c(N_c^2-1)}(\Pj_{SS}+\Pj_{PP})
                \mp \frac{1}{32N_c(N_c^2-1)}\Pj_{TT}  \nn \\
&& \Pj^{\pm}_5 \equiv \mp \frac{1}{32N_c(N_c^2-1)}(\Pj_{SS}+\Pj_{PP})
               + \ \frac{(2N_c\mp 1)}{96N_c(N_c^2-1)}\Pj_{TT} \; . \nn
\end{eqnarray}
They are obtained from the tree-level amputated Green's functions 
$\Lambda^{\pm (0)}_k$  imposing the following orthogonality conditions:
\be
\label{eq:orth}
\Tr\Pj^\pm_i \Lambda^{\pm (0)}_k = \delta_{ik} \qquad (i,k = 1,\dots,5)\; . 
\ee
The analogous projectors for the parity-violating operators can be found
in \cite{tassos}.

\section{One-loop results}\label{appd}
The one-loop expression for the quark propagator in a generic covariant gauge
is 
\be
S^{-1} = i\pslash \left\{1 + 
\frac{g_0^2}{16\pi^2}C_F\Big[\alpha L + B_\psi +\alpha\, d_\psi \Big]\right\},
\ee
and the amputated Green's functions of the bilinear operators in a generic 
covariant gauge are
\begin{eqnarray}
\Lambda_S  &= & S \Big\{ 1 + \frac{g_0^2}{16\pi^2} C_F
\Big[  -(3+\alpha ) L +B_S - \alpha\, (d_\psi - 1)\Big] \Big\} 
   \nonumber \\ 
\Lambda_V  &= & V \Big\{ 1 + \frac{g_0^2}{16\pi^2} C_F
\Big[ -\alpha\,  L + B_V - \alpha\, d_\psi \Big] \Big\} 
   -2 \frac{g_0^2}{16\pi^2} C_F \alpha 
\frac{p_\mu p \!\!\! / }{p^2} \nonumber \\
\Lambda_T  &= & T \Big\{ 1 + \frac{g_0^2}{16\pi^2} C_F 
\Big[ (1-\alpha ) L + B_T - \alpha\, (d_\psi+1) \Big]\Big\}  \; \\
\Lambda_A  &= & A \Big\{ 1 + \frac{g_0^2}{16\pi^2} C_F
\Big[ -\alpha\, L + B_A - \alpha\, d_\psi \Big] \Big\}
   -2 \frac{g_0^2}{16\pi^2} C_F\alpha 
\frac{p_\mu p \!\!\! / \gamma_5}{p^2} \nonumber \\
\Lambda_P  &= & P \Big\{ 1 + \frac{g_0^2}{16\pi^2} C_F
\Big[ -(3+\alpha ) L +B_P - \alpha\, (d_\psi - 1) \Big] \Big\} \; ,  
   \nonumber 
\end{eqnarray} 
with $L=\log(pa)^2$, and the $B_\Gamma$s are reported in Table~\ref{tab:bil}.
The terms proportional to the gauge parameter $\alpha$ can be given in 
simple form as functions of $d_\psi = \gamma_E -F_0 -1 = -4.792010$ 
(where $\gamma_E$ is the Euler constant and $F_0=4.3692252\cdots$).
They depend on the gluonic action, and they are independent of the $\rho$ 
parameter~\cite{stefano_neub}, as can be seen using gauge Ward Identities. 

The four-fermion operators that we have considered in the ``direct'' 
calculation has the form
\begin{eqnarray}
O_{\Gamma_A\Gamma_B} &=& \bar\psi_1 \Gamma_A \psi_2 \cdot  
                         \bar\psi_3 \Gamma_B \psi_4  \\
O_{\Gamma_A^c\Gamma_B^c} &=& \bar\psi_1 \Gamma_A t^C \psi_2 \cdot  
                             \bar\psi_3 \Gamma_B t^C \psi_4  \nonumber .
\end{eqnarray}
The one-loop expressions for the amputated Green's functions of the 
parity-conserving operators in a general covariant gauge are
\begin{eqnarray}
\Lambda_{S^cS^c}  &= & O_{S^cS^c} + \frac{g_0^2}{16\pi^2} \Bigg\{ 
    \Big[ -\frac{1}{2N_c} \Big(-2(3+\alpha) L +c_1 
    -2\alpha (d_\psi -1) \Big) 
    -N_c \Big( \alpha L +c_3 +\alpha \Big( d_\psi+\frac{1}{2} \Big) \Big) 
    \Big] O_{S^cS^c} \nonumber \\ 
&&  +\frac{N_c^2-1}{N_c^2} \Big( \Big( \frac{1}{2} L + c_2 \Big) O_{TT} 
    -\frac{1}{2} \alpha \, \sigma_{\rho \alpha} \otimes \sigma_{\rho \beta} \, 
    \frac{p_\alpha p_\beta}{p^2} \Big) 
    \nonumber \\
&&  
  + \frac{N_c^2-4}{N_c}  \Big( \Big( \frac{1}{2} L + c_2 \Big) O_{T^cT^c} 
    -\frac{1}{2} \alpha \, \sigma_{\rho \alpha} t^A \otimes 
      \sigma_{\rho \beta} t^A \,
    \frac{p_\alpha p_\beta}{p^2} \Big) 
    \Bigg\} \nonumber \\
\Lambda_{SS}  &= & O_{SS} + \frac{g_0^2}{16\pi^2} \Bigg\{
    \frac{N_c^2-1}{2N_c} \Big( -2(3+\alpha) L +c_1 
    -2\alpha (d_\psi -1) \Big) O_{SS} 
    \nonumber \\
&&  
 +4 \Big( \Big( \frac{1}{2} L + c_2 \Big) O_{T^cT^c} 
    -\frac{1}{2} \alpha \, \sigma_{\rho \alpha} t^A \otimes 
      \sigma_{\rho \beta} t^A \,
    \frac{p_\alpha p_\beta}{p^2} \Big) 
    \Bigg\} \nonumber \\ 
\Lambda_{P^cP^c}  &= &  O_{P^cP^c} + \frac{g_0^2}{16\pi^2} \Bigg\{
    \Big[ -\frac{1}{2N_c} \Big(-2(3+\alpha) L +c_1 
    -2\alpha (d_\psi -1)\Big) 
    -N_c \Big( \alpha L +c_3 +\alpha \Big( d_\psi+\frac{1}{2} \Big)
    \Big) \Big] O_{P^cP^c} \nonumber \\ 
&&  +\frac{N_c^2-1}{N_c^2} \Big( \Big( \frac{1}{2} L + c_2
    -\frac{1}{2} \alpha \Big) O_{TT}  
    +\frac{1}{2} \alpha \, \sigma_{\rho \alpha} \otimes \sigma_{\rho \beta} \,
    \frac{p_\alpha p_\beta}{p^2} \Big) 
    \nonumber \\
&&  
 + \frac{N_c^2-4}{N_c}  \Big( \Big( \frac{1}{2} L + c_2 
    -\frac{1}{2} \alpha  \Big) O_{T^cT^c}
    +\frac{1}{2} \alpha \, \sigma_{\rho \alpha} t^A \otimes 
      \sigma_{\rho \beta} t^A \, 
    \frac{p_\alpha p_\beta}{p^2} \Big)
    \Bigg\} \nonumber \\
\Lambda_{PP}  &= & O_{PP} + \frac{g_0^2}{16\pi^2} \Bigg\{
     \frac{N_c^2-1}{2N_c} \Big( -2(3+\alpha) L +c_1 
    -2\alpha (d_\psi -1)\Big) O_{PP} 
    \nonumber \\
&&  
 +4 \Big( \Big( \frac{1}{2} L + c_2 -\frac{1}{2} \alpha \Big) O_{T^cT^c} 
    +\frac{1}{2} \alpha \, \sigma_{\rho \alpha} t^A \otimes 
      \sigma_{\rho \beta} t^A \,
    \frac{p_\alpha p_\beta}{p^2} \Big) 
    \Bigg\} \nonumber \\ 
\Lambda_{V^cV^c}  &= & O_{V^cV^c} + \frac{g_0^2}{16\pi^2} \Bigg\{
    \Big[ -\frac{1}{2N_c} \Big(-2 \alpha L 
    +c_4 -2\alpha d_\psi \Big) 
    -N_c \Big( \Big( \frac{3}{2} +\alpha \Big) L +c_6 +\alpha d_\psi 
    \Big) \Big] O_{V^cV^c} 
    \nonumber \\
&&  
    +\Big( \frac{2}{N_c} +N_c \Big) \alpha \, \frac{p \!\!\! / t^A \otimes  
    p \!\!\! / t^A}{p^2}   \\ 
&&  +\frac{N_c^2-1}{N_c^2} \Big(\Big( \frac{3}{2} L +c_5 -\alpha \Big) O_{AA}  
    +\alpha \,\frac{p \!\!\! / \gamma_5 \otimes p \!\!\! / \gamma_5}{p^2}\Big) 
    \nonumber \\ 
&&  + \frac{N_c^2-4}{N_c} \Big(\Big(\frac{3}{2} L +c_5-\alpha \Big) O_{A^cA^c} 
  +\alpha \,\frac{p \!\!\! / \gamma_5 t^A \otimes p \!\!\! / t^A \gamma_5}{p^2}
    \Big) 
    \Bigg\} \nonumber \\
\Lambda_{VV}  &= & O_{VV} + \frac{g_0^2}{16\pi^2} \Bigg\{
     \frac{N_c^2-1}{2N_c} \Big( \Big( -2 \alpha L 
    +c_4 -2\alpha d_\psi \Big)  O_{VV} 
     -4\alpha \, \frac{p \!\!\! /  \otimes  p \!\!\! /}{p^2} \Big) \nonumber \\
&& +4 \Big( \Big( \frac{3}{2} L + c_5 -\alpha \Big) O_{A^cA^c} 
    + \alpha \, \frac{p \!\!\! / \gamma_5 t^A \otimes  
    p \!\!\! /  \gamma_5 t^A }{p^2} \Big) \Bigg\} \nonumber \\ 
\Lambda_{A^cA^c}  &= &  O_{A^cA^c} + \frac{g_0^2}{16\pi^2} \Bigg\{
    \Big[ -\frac{1}{2N_c} \Big(-2 \alpha L 
    +c_4 -2\alpha d_\psi \Big) 
    -N_c \Big( \Big( \frac{3}{2} +\alpha \Big) L +c_6 +\alpha d_\psi 
    \Big) \Big] O_{A^cA^c} 
    \nonumber \\
&&  
    +\Big(\frac{2}{N_c} +N_c \Big)\alpha\,\frac{p \!\!\! /\gamma_5 t^A \otimes 
    p \!\!\! / \gamma_5 t^A }{p^2}  
    \nonumber \\ 
&&  +\frac{N_c^2-1}{N_c^2} \Big(\Big( \frac{3}{2} L + c_5 -\alpha \Big) O_{VV} 
    +\alpha \, \frac{p \!\!\! / \otimes p \!\!\! / }{p^2} \Big) 
    \nonumber \\ 
&&  + \frac{N_c^2-4}{N_c}\Big(\Big( \frac{3}{2} L + c_5-\alpha \Big) O_{V^cV^c}
    +\alpha \, \frac{p \!\!\! / t^A \otimes p \!\!\! / t^A }{p^2} \Big) 
    \Bigg\} \nonumber \\
\Lambda_{AA}  &= & O_{AA} + \frac{g_0^2}{16\pi^2} \Bigg\{
   \frac{N_c^2-1}{2N_c} \Big( \Big( -2 \alpha L 
    +c_4 -2\alpha d_\psi \Big)  O_{AA} 
   -4\alpha \, \frac{p \!\!\! / \gamma_5  \otimes  p \!\!\! / \gamma_5 }{p^2}
    \Big) 
    \nonumber  \\
&& +4 \Big( \Big( \frac{3}{2} L + c_5-\alpha \Big) O_{V^cV^c} 
    + \alpha \, \frac{p \!\!\! / t^A \otimes p \!\!\! / t^A }{p^2} \Big) 
    \Bigg\} \nonumber \\ 
\Lambda_{T^cT^c}  &= &  O_{T^cT^c} + \frac{g_0^2}{16\pi^2} \Bigg\{
    \Big[ -\frac{1}{2N_c} \Big( 2(1-\alpha ) L 
    + c_7 -2\alpha (d_\psi +1) \Big) 
    -N_c \Big( (2+\alpha) L +c_9 +\alpha \Big( d_\psi -\frac{1}{2} \Big) 
    \Big) \Big] O_{T^cT^c} \nonumber \\ 
&&  +\frac{N_c^2-1}{N_c^2} \Big( 3 L + c_8 -\frac{3}{2} \alpha \Big) 
        (O_{SS}+O_{PP})  
    + \frac{N_c^2-4}{N_c}  \Big( 3 L + c_8 -\frac{3}{2} \alpha \Big) 
        (O_{S^cS^c}+O_{P^cP^c}) 
   \Bigg\} \nonumber \\
\Lambda_{TT}  &= &  O_{TT} + \frac{g_0^2}{16\pi^2} \Bigg\{
    \frac{N_c^2-1}{2N_c} \Big( 2(1-\alpha ) L 
    + c_7 -2\alpha (d_\psi +1) \Big) O_{TT} 
 +4 \Big( 3 L + c_8 -\frac{3}{2} \alpha \Big) (O_{S^cS^c}+O_{P^cP^c})\; 
  \Bigg\} . \nonumber 
\end{eqnarray}
\begin{table}[hbtp]
\begin{center}
  \begin{tabular}[btp]{|c||r|r|r|} \hline
          & $\rho=0.4$ & $\rho=1.0$ & $\rho=1.7$ \\ \hline
$c_1$   &   4.761193  &   8.939783  &  12.630887  \\ 
$c_2$   &  -0.142685  &  -0.488442  &  -0.790006  \\ 
$c_3$   &  -1.524484  &  -1.539241  &  -1.575409  \\ 
$c_4$   &   3.048969  &   3.078482  &   3.150818  \\ 
$c_5$   &  -0.428056  &  -1.465325  &  -2.370017  \\ 
$c_6$   &  -1.952540  &  -3.004566  &  -3.945426  \\ 
$c_7$   &   2.478227  &   1.124716  &  -0.009205  \\ 
$c_8$   &  -0.856112  &  -2.930650  &  -4.740034  \\ 
$c_9$   &  -2.095225  &  -3.493007  &  -4.735432  \\ 
\hline
  \end{tabular} 
\caption{Values of the finite constants in Landau gauge.}
\label{tab:qfdir}
\end{center}
\end{table}
The $c_i$ constants, which take the same values also for the mixings of 
the parity-violating operators (as we have explicitly verified), are reported 
in Table~\ref{tab:qfdir}. They correspond to the values of the finite parts 
in Landau gauge and, in terms of the finite parts of the bilinears, are:
\ba
c_1 = 2 B_S && \qquad c_2 = \frac{B_T-B_S}{8} 
\qquad \qquad c_3 = -\frac{B_S + 3 B_T + 4 B_V}{8}\nonumber\\
c_4 = 2 B_V && \qquad c_5 = \frac{B_V-B_S}{2} 
\qquad \qquad c_6 = -\frac{B_S+B_V}{2} \nonumber\\
c_7 = 2 B_T && \qquad c_8 = -\frac{3}{4}(B_S-B_T) 
\qquad c_9 = -\frac{1}{4} (3 B_S + B_T) .
\ea
The coefficients of the logarithms can also be re-expressed in terms of 
of the anomalous dimensions of the bilinears.

\end{appendix}
   
\end{document}